\documentclass[11pt,a4paper]{article}

\usepackage{jheppub}

\usepackage[english]{babel}

\allowdisplaybreaks[3]

\newcommand{\sect}[1]{section~\protect\ref{#1}}
\newcommand{\app}[1]{appendix~\protect\ref{#1}}

\newcommand{\vek}[1] {\boldsymbol{#1}}

\newcommand{\y}{\vek{y}}
\newcommand{\ytilde}{\tilde{\vek{y}}}

\newcommand{\dd}{\partial}

\newcommand{\half}{{\textstyle\frac{1}{2}}}

\newcommand{\ms}{\mskip 1.5mu}
\newcommand{\bs}{\mskip -1.5mu}

\subheader{\hfill DESY 13-037}

\title{Positivity bounds on double parton distributions}

\author{Markus Diehl}
\author{and Tomas Kasemets}
\affiliation{Deutsches Elektronen-Synchroton DESY, 22603 Hamburg,
  Germany} 

\emailAdd{markus.diehl@desy.de}
\emailAdd{tomas.kasemets@desy.de}

\abstract{Double hard scattering in proton-proton collisions is described
  in terms of double parton distributions.  We derive bounds on these
  distributions that follow from their interpretation as probability
  densities, taking into account all possible spin correlations between
  two partons in an unpolarized proton.  These bounds constrain the size
  of the polarized distributions and can for instance be used to set upper
  limits on the effects of spin correlations in double hard scattering.
  We investigate the stability of the bounds under leading-order DGLAP
  evolution to higher scales.}

\begin{document}

\maketitle

\section{Introduction}

In a time when the dynamics of the strong interaction in hadron-hadron
collisions is moving towards the domain of precision physics, there are
still aspects that are under poor theoretical and experimental control.
One of these aspects is double parton scattering, where two partons from
each proton have a hard interaction in a single proton-proton collision.
Correlations between the two hard interactions have been the subject of
several recent studies
\cite{Calucci:1997ii,Calucci:1999yz,DelFabbro:2000ds,%
  Flensburg:2011kj,Rogers:2009ke,Corke:2011yy,Domdey:2009bg,Rinaldi:2013vpa}.
The relevance of spin correlations in double parton scattering was pointed
out long ago \cite{Mekhfi:1985dv,Mekhfi:1983az} and recently followed up in
\cite{Diehl:2011tt,Diehl:2011yj}.  The studies in \cite{Manohar:2012jr}
and \cite{Kasemets:2012pr} have shown that spin correlations in the
production of two vector bosons by double hard scattering have observable
effects both on the interaction rate and on kinematic distributions.  Spin
correlations between the two partons are quantified by polarized double
parton distributions (DPDs), which describe for instance the difference of
the probability densities for finding two quarks with equal or with
opposite helicities.  It was argued in \cite{Diehl:2011yj} that such
correlations need not be small, and a recent study in the MIT bag model
\cite{Chang:2012nw} indeed found large spin correlations between quarks in
the valence region.  However, our knowledge of polarized DPDs is still
poor at best, and any information about them is of value.

In the present work, we derive model independent constraints on DPDs that
follow from their interpretation as probability densities for finding two
partons in a specified polarization state.  Similar positivity bounds have
been derived for single-parton distributions in the form of the Soffer
bound \cite{Soffer:1994ww} and of inequalities for transverse-momentum
dependent distributions \cite{Bacchetta:1999kz} and generalized parton
distributions \cite{Diehl:2005jf}.

The structure of this paper is as follows.  In the next section we set the
stage by introducing the DPDs for different polarizations and parton
species.  In \sect{sec:matrices} we derive the spin density matrices for
two partons inside an unpolarized proton, and in \sect{sec:bounds} we use
these matrices to derive bounds on polarized DPDs.  In
\sect{sec:evolution} we show that the homogeneous leading-order evolution
equations preserve these bounds when going to higher scales.  We conclude
in \sect{sec:sum} and give some technical details in two appendices.

%%%%%%%%%%%%%%%%%%%%%%%%%%%%%%%%%%%%%%%%%%%%%%%%%%%%%%%%%%%%%%

\section{Double parton distributions}
\label{sec:dpds}

Double parton distributions for quarks and antiquarks have been
extensively studied in \cite{Diehl:2011yj}, and we only review the
properties important for our purpose.  Since we will need a probability
interpretation, we restrict ourselves to distributions that are integrated
over the transverse parton momenta and that have a trivial color
structure.  In the parlance of \cite{Diehl:2011yj} these are collinear
color-singlet distributions.

Collinear DPDs depend on the longitudinal momentum fractions $x_1$ and
$x_2$ of the two partons and on the transverse distance $\y$ between them.
For two partons $a_1$ and $a_2$ in an unpolarized right-moving proton we
write
\begin{align}
  \label{eq:dpds}
 F_{a_1a_2}(x_1,x_2,\vek{y})
 & = 2p^+ (x_1\ms p^+)^{-n_1}\, (x_2\ms p^+)^{-n_2}
        \int \frac{dz^-_1}{2\pi}\, \frac{dz^-_2}{2\pi}\, dy^-\;
           e^{i\ms ( x_1^{} z_1^- + x_2^{} z_2^-)\ms p^+}
\nonumber \\
 & \quad \times \left<p|\, \mathcal{O}_{a_2}(0,z_2)\, 
            \mathcal{O}_{a_1}(y,z_1) \,|p\right> \,,
\end{align}
where $n_i = 1$ if parton number $i$ is a gluon and $n_i = 0$ otherwise.
We use light-cone coordinates $v^\pm = (v^0 \pm v^3) /\sqrt{2}$ and the
transverse component $\vek{v} = (v^1, v^2)$ for any four-vector $v$.  The
operators for quarks read
\begin{align}
\label{eq:quark-ops}
  \mathcal{O}_{a_i}(y,z_i)
   &= \bar{q}_i\bigl( y - \half z_i \bigr)\,
       \Gamma_{a_i} \, q_i\bigl( y + \half z_i \bigr)
   \Big|_{z_i^+ = y^+_{\phantom{i}} = 0,\; \vek{z}_i^{} = \vek{0}}
\end{align}
with projections
\begin{align}
  \label{eq:quark-proj}
  \Gamma_q & = \half \gamma^+ \,,
&
  \Gamma_{\Delta q} &= \half \gamma^+\gamma_5 \,,
&
  \Gamma_{\delta q}^j = \half i \sigma^{j+} \gamma_5 \quad (j=1,2)
\end{align}
onto unpolarized quarks ($q$), longitudinally polarized quarks ($\Delta
q$) and transversely polarized quarks ($\delta q$).  The field with
argument $y + \half z_i$ in $\mathcal{O}_{a_i}(y,z_i)$ is associated with
a quark in the amplitude of a double scattering process and the field with
argument $y - \half z_i$ with a quark in the complex conjugate amplitude.
The operators for gluons are
\begin{align}
\label{eq:gluon-ops}
  \mathcal{O}_{a_i}(y,z_i)
   &= \Pi_{a_i}^{jj'} \, G^{+j'}\bigl( y - \half z_i \bigr)\,
        G^{+j}\bigl( y + \half z_i \bigr)
   \Big|_{z_i^+ = y^+_{\phantom{i}} = 0,\; \vek{z}_i^{} = \vek{0}}
\end{align}
with projections
\begin{align}
  \label{eq:gluon-proj}
  \Pi_g^{jj'}  &= \delta^{jj'} \,,
&
  \Pi_{\Delta g}^{jj'} &= i\epsilon^{jj'} \,,
&
  [\Pi_{\delta g}^{kk'}]^{jj'} &= \tau^{jj'\!,kk'}
\end{align}
onto unpolarized gluons ($g$), longitudinally polarized gluons ($\Delta
g$) and linearly polarized gluons ($\delta g$). The tensor
\begin{align}
  \tau^{jj'\!,kk'} = \half \ms \bigl( \delta^{jk}\delta^{j'k'} 
     + \delta^{jk'}\delta^{j'k} - \delta^{jj'}\delta^{kk'} \bigr)
\end{align}
satisfies $\tau^{jj'\!,kk'} \tau^{kk'\!,\,ll'} = \tau^{jj'\!,\,ll'}$ and
is symmetric and traceless in each of the index pairs $(jj')$ and $(kk')$.
Note that for gluons $\delta g$ denotes linear polarization, i.e.\ the
interference between gluons whose helicities differ by two units in the
scattering amplitude and its conjugate, while for quarks $\delta q$
symbolizes transverse polarization, where the interference is between
quarks with a helicity difference of one unit.
Since we limit ourselves to color-singlet distributions, a sum over the
color indices of the quark fields in \eqref{eq:quark-ops} and the gluon
fields in \eqref{eq:gluon-ops} is implied.  We do not write out the Wilson
lines that make the operators gauge invariant.

The different spin projections lead to a large number of DPDs.  For
collinear color-singlet distributions, several polarization combinations
are zero due to time reversal and parity invariance.  This concerns the
DPDs with one longitudinally polarized and one unpolarized parton, as well
as those with one longitudinally polarized parton and one transversely
polarized (anti)quark or linearly polarized gluon.  A decomposition of the
nonzero distributions for two quarks in terms of real-valued scalar
functions has already been given in \cite{Diehl:2011yj}:
\begin{align}
\label{eq:def-qq}
  F_{qq}(x_1,x_2,\y) & = f_{qq}(x_1,x_2,y) \,,
  \nonumber\\
  F_{\Delta q \Delta q}(x_1,x_2,\y) & =
    f_{\Delta q \Delta q}(x_1,x_2,y) \,,
  \nonumber\\
  F_{q \ms \delta q}^j(x_1,x_2,\y) & =
    \tilde{\vek{y}}^j M f_{q \ms \delta q}(x_1,x_2,y) \,,
  \nonumber\\
  F_{\delta q \ms q}^j(x_1,x_2,\y) & =
    \tilde{\vek{y}}^j M f_{\delta q \ms q}(x_1,x_2,y) \,,
  \nonumber\\
  F_{\delta q \delta q}^{jj'}(x_1,x_2,\y) & =
    \delta^{jj'} f_{\delta q \delta q}(x_1,x_2,y)
    +  2\tau^{jj'\!,\y\y}M^2 f_{\delta q \delta q}^t(x_1,x_2,y) \,,
\end{align}
where $M$ is the proton mass, $\ytilde^j = \epsilon^{jj'} \y^{j'}$ and
$y=\sqrt{\y^2}$.  We use a shorthand notation where vectors $\y$ or
$\ytilde$ appearing as an index of $\tau$ denote contraction, i.e.\
$\tau^{jj'\!,\y\y} = \tau^{jj'\!,kk'}\, \y^k \y^{k'}$ etc.  Decompositions
analogous to \eqref{eq:def-qq} hold for quark-antiquark distributions and
for the distributions of two antiquarks.

Since quarks and gluons mix under evolution, we also need to consider DPDs
involving gluons.  We define
\begin{align}
\label{eq:def-qg}
  F_{qg}(x_1,x_2,\y) & = f_{qg}(x_1,x_2,y) \,,
  \nonumber\\
  F_{\Delta q \Delta g}(x_1,x_2,\y) & =
     f_{\Delta q \Delta g}(x_1,x_2,y) \,,
  \nonumber\\
  F_{q \ms \delta g}^{jj'}(x_1,x_2,\y) & =
     \tau^{jj'\!,\y\y} M^2 f_{q \ms \delta g}(x_1,x_2,y) \,,
  \nonumber\\
  F_{\delta q \ms g}^j(x_1,x_2,\y) & =
     \ytilde^j M f_{\delta q \ms g}(x_1,x_2,y) \,,
  \nonumber\\
  F_{\delta q \delta g}^{j,kk'}(x_1,x_2,\y) & =
     \!\! {}- \tau^{\ytilde j, kk'} M f_{\delta q \delta g}(x_1,x_2,y)
  \nonumber\\
  & \quad - \bigl( \ytilde^j \tau^{kk'\!,\y\y}
                 + \y^j \tau^{kk'\!,\y\ytilde} \bigr) \,
              M^3 f_{\delta q \delta g}^t(x_1,x_2,y)
\end{align}
for quark-gluon distributions, with analogous expressions for gluon-quark
distributions and distributions where the quark is replaced by an
antiquark.  For two-gluon distributions we write
\begin{align}
\label{eq:def-gg}
  F_{gg}(x_1,x_2,\y) & = f_{gg}(x_1,x_2,y) \,,
  \nonumber\\
  F_{\Delta g \Delta g}(x_1,x_2,\y) & =
    f_{\Delta g \Delta g}(x_1,x_2,y) \,,
  \nonumber\\
  F_{g \ms \delta g}^{jj'}(x_1,x_2,\y) & =
    \tau^{jj'\!,\y\y} M^2 f_{g \ms \delta g}(x_1,x_2,y) \,,
  \nonumber\\
  F_{\delta g \ms g}^{jj'}(x_1,x_2,\y) & =
    \tau^{jj'\!,\y\y} M^2 f_{\delta g \ms g}(x_1,x_2,y) \,,
  \nonumber\\
  F_{\delta g \delta g}^{jj',kk'}(x_1,x_2,\y) & =
    \half\ms \tau^{jj'\!,\,kk'} f_{\delta g \delta g}(x_1,x_2,y) \,,
  \nonumber\\                           
  & \quad + \bigl( \tau^{jj'\!,\y\ytilde} \tau^{kk'\!,\y\ytilde}
      - \tau^{jj'\!,\y\y} \tau^{kk'\!,\y\y} \bigr) \,
      M^4 f_{\delta g \delta g}^t(x_1,x_2,y) \,.
\end{align}
We remark that, although linear gluon polarization is described by a
tensor with two indices, the restriction that this tensor is symmetric and
traceless gives rise to the same number of distributions as for transverse
quark polarization, which is described by a vector.  The prefactors in
\eqref{eq:def-qg} and \eqref{eq:def-gg} have been chosen such that we will
obtain a simple correspondence between quark and gluon distributions in
the spin density matrices to be derived in the next section.

Note that DPDs involving gluons are not only relevant in the context of
evolution but also enter directly in important double scattering processes
such as the production of jets.  Their properties are hence of
considerable practical interest.

In complete analogy to the case of collinear single-parton distributions,
the DPDs we have introduced can be interpreted as probability densities
for finding two partons inside an unpolarized proton, with a relative
transverse distance $\vek{y}$ and with longitudinal momentum fractions
$x_1$ and $x_2$.  This becomes evident from their appearance in the cross
section formulae for double parton scattering \cite{Diehl:2011yj}.  It can
also be seen from a representation in terms of parton creation and
annihilation operators or from a representation in terms of the light-cone
wave functions of the proton, which are straightforward extensions of the
corresponding representations for single-parton distributions (given for
instance in sections 3.4 and 3.11 of \cite{Diehl:2003ny}).

As in the case of single-parton densities, this interpretation does
however not strictly hold in QCD, because the distributions are defined
with subtractions from the ultraviolet region of parton momenta.  The
subtraction terms can in principle invalidate the positivity of the
distributions.  Nevertheless, it is useful to explore the consequences of
the probability interpretation as a guide for developing physically
intuitive models of the distributions.  This holds in particular if one
works in leading order of $\alpha_s$, where the connection between parton
distributions and physical cross sections (which must of course be
positive semi-definite) is most direct.

%%%%%%%%%%%%%%%%%%%%%%%%%%%%%%%%%%%%%%%%%%%%%%%%%%%%%%%%%%%%%%

\section{Two-parton spin density matrices}
\label{sec:matrices}

The polarization state of two partons in an unpolarized proton is
described by a spin density matrix that can be written in terms of the
DPDs we introduced in the previous section.  We start by trading the
projection operators \eqref{eq:quark-proj} and \eqref{eq:gluon-proj} for
operators that project onto quarks or gluons of definite helicity.  We can
then easily write down the spin density matrix for two partons in the
helicity basis.

The projection operators $\Gamma_{\lambda' \lambda}$ for quarks, where
$\lambda$ ($\lambda'$) refers to the quark helicity in the amplitude
(conjugate amplitude), are given by
\begin{align}
  \Gamma_{++} & = \frac{\gamma^+}{4}\, (1 + \gamma_5)
                = \frac{\Gamma_q + \Gamma_{\Delta q}}{2} \,,
&
  \Gamma_{+-} & = \phantom{-}\frac{i\sigma^{+1}}{4}\, (1 - \gamma_5)
                = \frac{\Gamma_{\delta q}^1 + i\Gamma_{\delta q}^2}{2} \,,
\nonumber\\
  \Gamma_{--} & = \frac{\gamma^+}{4}\, (1 - \gamma_5)
                = \frac{\Gamma_q - \Gamma_{\Delta q}}{2} \,,
&
  \Gamma_{-+} & = -\frac{i\sigma^{+1}}{4}\, (1 + \gamma_5) 
                = \frac{\Gamma_{\delta q}^1 - i\Gamma_{\delta q}^2}{2} \,.
\end{align}
Here we use the phase conventions for spin-half fields specified in
\cite{Diehl:2001pm}.  The projection operators $\Pi_{\lambda'
  \lambda}^{jj'}$ for gluons, where $\lambda$ and $j$ ($\lambda'$ and
$j'$) refer to the amplitude (conjugate amplitude), can be constructed
from the polarization vectors
\begin{align}
  \vek{\epsilon}_+ &= - \frac{1}{\sqrt{2}}\, \bigl(\ms 1, i \ms\bigr) \,,
&
  \vek{\epsilon}_-  &= \frac{1}{\sqrt{2}}\, \bigl(\ms 1, -i \ms\bigr)
\end{align}
and read
\begin{align}
  \Pi_{++}^{jj'} = \bigl( \epsilon_+^{j} \bigr)^*\, \epsilon_+^{j'}
               & = \frac{1}{2} \left( \Pi_g^{jj'} + \Pi_{\Delta g}^{jj'} 
                               \right) \,,
\nonumber\\
  \Pi_{--}^{jj'} = \bigl( \epsilon_-^{j} \bigr)^*\, \epsilon_-^{j'}
               & = \frac{1}{2} \left( \Pi_g^{jj'} - \Pi_{\Delta g}^{jj'}
                               \right) \,,
\nonumber \\
  \Pi_{+-}^{jj'} = \bigl( \epsilon_-^{j} \bigr)^*\, \epsilon_+^{j'}
               & =   - \bigl[ \Pi_{\delta g}^{11} \bigr]^{jj'} 
                   - i \bigl[ \Pi_{\delta g}^{12} \bigr]^{jj'} \,,
  \phantom{\frac{1}{1}}
\nonumber \\
  \Pi_{-+}^{jj'} = \bigl( \epsilon_+^{j} \bigr)^*\, \epsilon_-^{j'}
               & =   - \bigl[ \Pi_{\delta g}^{11} \bigr]^{jj'} 
                   + i \bigl[ \Pi_{\delta g}^{12} \bigr]^{jj'} \,.
  \phantom{\frac{1}{1}}
\end{align}
We can now organize the distributions in matrices where the columns (rows)
correspond to helicity states $++,-+,+-,--$ of the two partons in the
amplitude (conjugate amplitude).  The spin density matrix for two quarks
reads
\begin{align}
  \label{eq:qq-matrix}
\rho &= \frac{1}{4} \begin{pmatrix}
    f_{qq} + f_{\Delta q\Delta q}
  & -ie^{i\varphi_y} y M f_{\delta q \ms q}
  & -ie^{i\varphi_y} y M f_{q \ms \delta q}
  & 2e^{2i\varphi_y} y^2 M^2 f_{\delta q\delta q}^t
\\[0.2cm]
    ie^{-i\varphi_y} y M f_{\delta q \ms q}
  & f_{qq} - f_{\Delta q\Delta q}
  & 2f_{\delta q \delta q}
  & -ie^{i\varphi_y} y M f_{q \ms \delta q} 
\\[0.2cm]
    ie^{-i\varphi_y} y M f_{q \ms \delta q}
  & 2f_{\delta q \delta q}
  & f_{qq} - f_{\Delta q\Delta q}
  & -ie^{i\varphi_y} y M f_{\delta q \ms q} 
\\[0.2cm]
    2e^{-2i\varphi_y} y^2 M^2 f_{\delta q \delta q}^t
  & ie^{-i\varphi_y} y M f_{q \ms \delta q}
  & ie^{-i\varphi_y} y M f_{\delta q \ms q}
  & f_{qq} + f_{\Delta q \Delta q}
\end{pmatrix} \,,
\end{align}
where the angle $\varphi_y$ describes the orientation of the vector $\y =
y\, (\cos\varphi_y, \sin\varphi_y)$ in the transverse plane.  With the
caveat spelled out at the end of the previous section, the diagonal matrix
elements can be interpreted as the probability densities for finding two
partons in definite helicity states inside an unpolarized proton.
Specifically, $f_{qq}+f_{\Delta q\Delta q}$ is the probability density for
finding two quarks with positive helicities, which in an unpolarized
proton is equal to the probability density for finding two quarks with
negative helicities.  The probability density for finding two quarks with
opposite helicities is $f_{qq}-f_{\Delta q \Delta q}$.
The off-diagonal elements of $\rho$ describe helicity interference, with
$f_{\delta q \delta q}^t$ in the right upper corner corresponding for
instance to the case where both quarks have negative helicity in the
amplitude and positive helicity in the conjugate amplitude.  This leads to
a helicity difference between the amplitude and its conjugate, which is
balanced by two units of orbital angular momentum indicated by an
exponential $e^{2i\varphi_{y}}$ and an associated factor $y^2$.  By
contrast, $f_{\delta q \delta q}$ describes the case when the helicity
difference is $+1$ for one quark and $-1$ for the other, so that the
overall helicity is balanced.

Turning now to gluons, we have a spin density matrix
\begin{align}
  \label{eq:qg-matrix}
\frac{1}{4} \begin{pmatrix}
     f_{qg} + f_{\Delta q\Delta g} & 
      -ie^{i\varphi_y} y M f_{\delta q \ms g} & 
      -e^{2i\varphi_y} y^2 M^2 f_{q \ms \delta g} & 
      -2ie^{3i\varphi_y} y^3 M^3 f_{\delta q\delta g}^t
    \\[0.2cm]
     ie^{-i\varphi_y} y M f_{\delta q \ms g} & 
      f_{qg} - f_{\Delta q\Delta g} & 
      -2i e^{i\varphi_y} y M f_{\delta q \delta g} & 
      -e^{2i\varphi_y} y^2 M^2 f_{q \ms \delta g} 
    \\[0.2cm]
     -e^{-2i\varphi_y} y^2 M^2 f_{q \ms \delta g} & 
      2ie^{-i\varphi_y} y M f_{\delta q \delta g} & 
      f_{qg} - f_{\Delta q\Delta g} & 
      -ie^{i\varphi_y} y M f_{\delta q \ms g} 
    \\[0.2cm]
      2ie^{-3i\varphi_y} y^3 M^3 f_{\delta q \delta g}^t & 
      -e^{-2i\varphi_y} y^2 M^2 f_{q \ms \delta g} &
      ie^{-i\varphi_y} y M f_{\delta q\ms g} &
      f_{qg} + f_{\Delta q \Delta g}
  \end{pmatrix}
\end{align}
for quark-gluon distributions and an analogous matrix for gluon-quark
distributions.  For two-gluon distributions we find
\begin{align}
  \label{eq:gg-matrix}
\frac{1}{4} \begin{pmatrix}
     f_{gg} + f_{\Delta g\Delta g} & 
      -e^{2i\varphi_y} y^2 M^2 f_{\delta g \ms g} & 
      -e^{2i\varphi_y} y^2 M^2 f_{g \ms \delta g} & 
      -2e^{4i\varphi_y} y^4 M^4 f_{\delta g\delta g}^t
    \\[0.2cm]
      -e^{-2i\varphi_y} y^2 M^2 f_{\delta g \ms g} & 
      f_{gg} - f_{\Delta g\Delta g} & 
      2f_{\delta g \delta g} & 
      -e^{2i\varphi_y} y^2 M^2 f_{g \ms \delta g} 
    \\[0.2cm]
      -e^{-2i\varphi_y} y^2 M^2 f_{g \ms \delta g} & 
      2f_{\delta g \delta g} & 
      f_{gg} - f_{\Delta g\Delta g} & 
      -e^{2i\varphi_y} y^2 M^2 f_{\delta g \ms g} 
    \\[0.2cm]
      -2e^{-4i\varphi_y} y^4 M^4 f_{\delta g \delta g}^t & 
      -e^{-2i\varphi_y} y^2 M^2 f_{g \ms \delta g} &
      -e^{-2i\varphi_y} y^2 M^2 f_{\delta g \ms g} &
      f_{gg} + f_{\Delta g \Delta g}
  \end{pmatrix} \,.
\end{align}
The matrices for distributions where quarks are replaced by antiquarks are
analogous to \eqref{eq:qq-matrix} and \eqref{eq:qg-matrix}.  We see that
the parameterization of DPDs in the previous section gives simple
expressions for the spin density matrices and similar structures for all
types of partons.

The difference in spin between quarks and gluons causes the different
dependence on the azimuthal angle $\varphi_y$ in \eqref{eq:qq-matrix},
\eqref{eq:qg-matrix} and \eqref{eq:gg-matrix}.  A mismatch of $n$ units
between the sum of parton helicities in the amplitude and its conjugate
goes along with an exponential $e^{\pm n i \varphi_y}$ and an associated
factor $y^n$.

%%%%%%%%%%%%%%%%%%%%%%%%%%%%%%%%%%%%%%%%%%%%%%%%%%%%%%%%%%%%%% 

\section{Positivity bounds} 
\label{sec:bounds}

We now show how the probability interpretation of DPDs constrains the size
of the polarized distributions.
Since the probability density for finding two partons in a general
polarization state is positive semi-definite, we have
\begin{equation}
  \label{eq:pos-def-matrix}
\sum_{\lambda_1'\lambda_2'\ms \lambda_1^{}\lambda_2^{}}
 v^*_{\lambda_1'\lambda_2'} \,
 \rho_{(\lambda_1'\lambda_2')(\lambda_1^{}\lambda_2^{})}^{\phantom{*}} \,
 v_{\lambda_1^{}\lambda_2^{\phantom{\prime}}}^{\phantom{*}}
 \,\geq\, 0
\end{equation}
with arbitrary complex coefficients $v_{\lambda_1\lambda_2}$ normalized as
$\sum_{\lambda_1\lambda_2}|v_{\lambda_1\lambda_2}|^2 = 1$. The helicity
matrices are therefore positive semi-definite.  The same property has been
derived for the spin density matrices associated with transverse-momentum
dependent distributions \cite{Bacchetta:1999kz} or generalized parton
distributions \cite{Diehl:2005jf}.

To simplify the algebra, we first cast all helicity matrices into a common
form that is independent of the angle $\varphi_y$.  This is achieved by
unitary transformations, multiplying by a matrix $U$ from the right and by
$U^\dagger$ from the left.  The transformation matrices for the parton
combinations in \eqref{eq:qq-matrix} to \eqref{eq:gg-matrix} are
\begin{align}
  U_{qq} & = \operatorname{diag} \bigl( -e^{2i\varphi_y},
             -ie^{i\varphi_y}, -ie^{i\varphi_y}, 1 \,\bigr) \,,
  \nonumber\\
  U_{qg} & = \operatorname{diag} \bigl( \, ie^{3i\varphi_y},
             -e^{2i\varphi_y}, -ie^{i\varphi_y}, 1 \,\bigr) \,,
  \nonumber\\
  U_{gg} & = \operatorname{diag} \bigl( \, e^{4i\varphi_y},
             -e^{2i\varphi_y}, -e^{2i\varphi_y}, 1 \,\bigr) \,.
\end{align}
After these transformations and their analog for gluon-quark
distributions, the spin density matrices can be written as
\begin{align}
\label{eq:mat}
\rho & = \frac{1}{4} \begin{pmatrix}
   f_{ab} + f_{\Delta a \Delta b} & h_{\delta a \ms b}       &
   h_{a \ms \delta b}       & -2h_{\delta a \delta b}^t      \\
   h_{\delta a \ms b}       & f_{ab} - f_{\Delta a \Delta b} &
   2h_{\delta a \delta b}   &  h_{a \ms \delta b}            \\
   h_{a \ms \delta b}       & 2h_{\delta a \delta b}         &
   f_{ab} - f_{\Delta a \Delta b} & h_{\delta a \ms b}       \\
   -2h_{\delta a \delta b}^t     &  h_{a \ms \delta b}       &
   h_{\delta a \ms b}       & f_{ab} + f_{\Delta a \Delta b} 
 \end{pmatrix}
\end{align}
with the following identification of distributions for different parton
combinations:
\begin{align}
  \label{dist-list}
  f_{ab} & = f_{qq} \,,\; f_{qg}\,,\; f_{gq} \,,\; f_{gg} \,,
\nonumber \\[0.2em]
  f_{\Delta a \Delta b} & = f_{\Delta q \Delta q} \,,\;
  f_{\Delta q \Delta g} \,,\;
  f_{\Delta g \Delta q} \,,\;
  f_{\Delta g \Delta g} \,,
\nonumber \\[0.2em]
  h_{\delta a \ms b} & = y M f_{\delta q \ms q} \,,\;
  y M f_{\delta q \ms g} \,,\;
  y^2 M^2 f_{\delta g \ms q} \,,\;
  y^2 M^2 f_{\delta g \ms g} \,,
\nonumber \\[0.2em]
  h_{a \ms \delta b} & = y M f_{q \ms \delta q} \,,\;
  y^2 M^2 f_{q \ms \delta g} \,,\;
  y M f_{g \ms \delta q} \,,\;
  y^2 M^2 f_{g \ms \delta g} \,,
\nonumber \\[0.2em]
  h_{\delta a \delta b} & = f_{\delta q \delta q} \,,\;
  y M f_{\delta q \delta g} \,,\;
  y M f_{\delta g \delta q} \,,\;
  f_{\delta g \delta g} \,,
\nonumber \\[0.2em]
  h_{\delta a \delta b}^t & = y^2 M^2 f_{\delta q \delta q}^t \,,\;
  y^3 M^3 f_{\delta q \delta g}^t \,,\;
  y^3 M^3 f_{\delta g \delta q}^t \,,\;
  y^4 M^4 f_{\delta g \delta g}^t \,.
\end{align}
Analogous expressions hold if quarks are replaced by antiquarks.
Positivity\footnote{%
  For ease of language we use ``positivity'' in the sense of ``positive
  semi-definite'' here and in the following.}
of the diagonal elements of $\rho$ yields the trivial bounds
\begin{align}
  \label{first-ineq}
  f_{ab} \,\geq\, \big| f_{\Delta a \Delta b} \big| \,.
\end{align}
The principal minors of the two-dimensional sub-spaces must be positive
semi-definite as well, which gives upper bounds on the distributions for
one or two transversely or linearly polarized partons:
\begin{align}
  \label{second-ineq}
  f_{ab} + f_{\Delta a \Delta b} & \,\geq\,
    2\ms \big| h_{\delta a \delta b}^t \big| \,,
\nonumber\\
  f_{ab} - f_{\Delta a \Delta b} & \,\geq\,
    2\ms \big| h_{\delta a \delta b} \big| \,,
\nonumber\\
  f_{ab}^2 \geq (f_{ab} + f_{\Delta a \Delta b}) & 
    (f_{ab} - f_{\Delta a \Delta b}) \,\geq\, h_{\delta a \ms b}^2 \,,
\nonumber\\
  f_{ab}^2 \geq (f_{ab} + f_{\Delta a \Delta b}) &
    (f_{ab} - f_{\Delta a \Delta b}) \,\geq\, h_{a \ms \delta b}^2 \,.
\end{align}
The principal minors of dimension three, as well as $\det (\rho)$ provide
further bounds, which are rather cumbersome and will not be given here.
The strongest bounds can be obtained from the positivity of the
eigenvalues of $\rho$, which is a sufficient and necessary condition for
the positivity of $\rho$.  Calculating the eigenvalues we obtain
\begin{align}
   f_{ab} + h_{\delta a \delta b} - h_{\delta a \delta b}^t   
      & \pm \sqrt{(h_{\delta a \ms b} +  h_{a \ms \delta b})^2 +
        (f_{\Delta a \Delta b} - h_{\delta a \delta b}
       - h_{\delta a \delta b}^t)^2}  \,\geq\, 0 \,,
  \nonumber\\
  f_{ab} - h_{\delta a \delta b} + h_{\delta a \delta b}^t 
      & \pm \sqrt{(h_{\delta a \ms b} -  h_{a \ms \delta b})^2 +
        (f_{\Delta a \Delta b} + h_{\delta a \delta b}
       + h_{\delta a \delta b}^t)^2}  \,\geq\, 0 \,.
\end{align}
These inequalities set upper limits on the size of spin correlations
between two partons in an unpolarized proton. They can be used either to
construct double parton distributions or to put limits on polarization
effects in double hard scattering processes.

We note that positive semidefinite combinations of DPDs were discussed
already in the pioneering studies \cite{Mekhfi:1985dv,Mekhfi:1983az}.
Distributions that involve a helicity mismatch between the amplitude and
its conjugate (see \sect{sec:matrices}) were however not considered in
that work.  The derivation in \cite{Mekhfi:1985dv,Mekhfi:1983az} thus
corresponds to our results \eqref{first-ineq} and \eqref{second-ineq} if
all distributions multiplied with a power of $y$ in \eqref{dist-list} are
set to zero.

%%%%%%%%%%%%%%%%%%%%%%%%%%%%%%%%%%%%%%%%%%%%%%%%%%%%%%%%%%%%%% 

\section{Stability under evolution}
\label{sec:evolution}

The ultraviolet subtractions mentioned at the end of \sect{sec:dpds}
induce a scale dependence, which for collinear single-parton distributions
is described by the DGLAP evolution equations.  While the subtractions
themselves may invalidate positivity of the distributions and thus their
density interpretation, the evolution equations can be interpreted in a
probabilistic manner provided that one takes the leading-order
approximation of the evolution kernels
\cite{Durand:1986te,Collins:1988wj}.  Specifically, one finds that if
parton distributions are positive semi-definite at a certain scale, this
property is preserved by leading-order evolution to higher scales.  This
also holds for the Soffer inequality, which expresses positivity in the
sector of transverse quark polarization
\cite{Barone:1997fh,Bourrely:1997bx}.
For evolution at next-to-leading order in $\alpha_s$ the situation is less
clear-cut and a discussion of positivity depends in particular on the
scheme in which the distributions are defined.  We refer to
\cite{Altarelli:1998gn} and \cite{Vogelsang:1997ak,Martin:1997rz} for a
discussion of the situation for longitudinal and transverse parton
polarization, respectively.

Returning to double parton distributions, we now show that the bounds
derived in the previous section are stable under leading-order evolution
to higher scales.  The strategy for the derivation is as follows: we first
introduce linear combinations of double parton distributions whose
positivity is necessary and sufficient for the positivity of the spin
density matrices and then show that these linear combinations remain
positive semi-definite under evolution.  The positivity of the spin
density matrices then guarantees the stability of the positivity bounds.

%%%%%%%%%%%%%%

\subsection{Evolution of double parton distributions}

To begin with, let us specify the evolution of collinear DPDs in the
color-singlet sector.  We use the homogeneous evolution equations in the
transverse position representation, see e.g.\ equation (5.93) in
\cite{Diehl:2011yj}.  These equations apply at nonzero $\vek{y}$ if
$F_{a_1a_2}(x_1,x_2,\vek{y})$ is defined via \eqref{eq:dpds} with the
operators $\mathcal{O}_{a_1}(y,z_1)$ and $\mathcal{O}_{a_2}(0,z_2)$
renormalized by standard $\overline{\text{MS}}$ subtraction.  The
inhomogeneous term for the splitting of one parton into two that has been
previously considered in the literature
\cite{Kirschner:1979im,Shelest:1982dg,Snigirev:2003cq,%
  Gaunt:2009re,Ceccopieri:2010kg} does not appear in this case.  As
discussed in \cite{Diehl:2011tt,Diehl:2011yj}, a consistent formulation of
factorization for double parton scattering does not yet exist, so that it
remains unclear how DPDs should best be defined (and how they evolve).
For simplicity we will limit our present investigation to the homogeneous
evolution equations.

It is useful for our purpose to take different renormalization scales
$\mu_1$ and $\mu_2$ for the two partons, corresponding to separate
ultraviolet renormalization of $\mathcal{O}_{a_1}$ and $\mathcal{O}_{a_2}$
in \eqref{eq:dpds}.  The evolution equation for the unpolarized double
quark distributions in the first scale then reads
\begin{align}
  \label{eq:evol-parton-1}
  \frac{\dd f_{qq}(x_1,x_2,y; \mu_1,\mu_2)}{\dd\tau_1}
    & = P_{qq} \otimes_1 f_{qq} + P_{qg} \otimes_1 f_{gq} \,,
\end{align}
where
\begin{align}
\label{eq:otim}
    P_{ab}( \, .\, ) \otimes_1
       f_{bc}( \, .\, ,x_2,y;\mu_1,\mu_2) 
    & = \int_{x_1}^{1-x_2} \frac{d u_1}{u_1}
        P_{ab}\left( \frac{x_1}{u_1} \right)
          f_{bc}(u_1,x_2,y;\mu_1,\mu_2)
\end{align}
is a convolution in the first argument of the DPDs with the leading-order
splitting functions $P_{ab}$ known from DGLAP evolution of single-parton
distributions.  We note that the leading-order splitting functions are the
same for quarks and antiquarks, i.e.\ one has $P_{qq} =
P_{\bar{q}\bar{q}}$, $P_{qg} = P_{\bar{q}g}$, $P_{gq} = P_{g\bar{q}}$ and
analogous relations for polarized partons.  In \app{ap:evo} we give the
explicit evolution equations for all polarized DPDs and list the
associated splitting functions.

The evolution variable in \eqref{eq:evol-parton-1} is taken as
\begin{align}
  \tau_1 = \int^{\mu_1^2} \frac{d\mu^2}{\mu^2}\,
     \frac{\alpha_s(\mu)}{2\pi} \,,
\end{align}
where the lower limit of integration is irrelevant in the derivative $\dd
f/\dd\tau_1$.  The use of $\tau_1$ is just a matter of convenience as it
removes the running coupling from the leading-order splitting functions.

The analog of \eqref{eq:evol-parton-1} for the scale associated with the
second parton is
\begin{align}
  \label{eq:evol-parton-2}
  \frac{\dd f_{qq}(x_1,x_2,y; \mu_1,\mu_2)}{\dd\tau_2}
    & = P_{qq} \otimes_2 f_{qq} + P_{qg} \otimes_2 f_{qg} \,.
\end{align}
The evolution equation for equal scales, i.e.\ for $f_{qq}(x_1,x_2,y;
\mu,\mu)$, is readily obtained by adding the right-hand sides of
\eqref{eq:evol-parton-1} and \eqref{eq:evol-parton-2}.  We will show that
positivity is preserved for separate evolution in $\mu_1$.  The same then
obviously holds for evolution in $\mu_2$ and hence for the evolution in a
single common scale $\mu_1 = \mu_2$.

%%%%%%%%%%%%%%

\subsection{Linear combinations of DPDs}

A key ingredient in our argument is to form suitable linear combinations
of double parton distributions, which we now introduce.
Positivity of the spin density matrix $\rho$ means that $v^\dagger
\bs\rho\, v \geq 0$ for any complex vector $v$, as we spelled out in
\eqref{eq:pos-def-matrix}.  Parameterizing the vector as
\begin{align}
  v^T = (a_1 + ib_1, a_2 + ib_2, a_3 + ib_3, a_4 + ib_4)
\end{align}
with real numbers $a_i$, $b_i$ and performing the multiplication with the
matrix in \eqref{eq:mat} gives
\begin{align}
  \label{eq:cond1}
Q_{ab}^+ & = c_1 f_{ab} + c_2\ms h_{a \ms \delta b}
    + c_3\ms f_{\Delta a \Delta b} + c_4\ms h_{\delta a \ms b}
    + c_5\ms h_{\delta a \delta b} + c_6\ms h_{\delta a \delta b}^t
  \,\geq\, 0 \,,
\end{align}
where $Q_{ab}^+ = 4 v^\dagger \bs\rho\, v$ and the coefficients $c_i$ are
given by
\begin{align}
  c_1 & =
    a_1^2 + b_1^2 + a_2^2 + b_2^2 + a_3^2 + b_3^2 + a_4^2 + b_4^2 \,,
&
  c_2 & = 2 ( a_1 a_3 + b_1 b_3 + a_2 a_4 + b_2 b_4 ) \,,
\nonumber\\
  c_3 & =
   a_1^2 + b_1^2 - a_2^2 - b_2^2 - a_3^2 - b_3^2 + a_4^2 + b_4^2 \,,
&
  c_4 & = 2 ( a_1 a_2 + b_1 b_2 + a_3 a_4 + b_3 b_4 ) \,,
\nonumber\\
  c_5 & = 4 ( a_2 a_3 + b_2 b_3 ) \,,
&
  c_6 & = -4 ( a_1 a_4 + b_1 b_4 ) \,.
\end{align}
We will prove the stability of the positivity bounds by showing that for
arbitrary values of $a_i$ and $b_i$ the inequality \eqref{eq:cond1} is
stable under evolution to higher scales.  It will be convenient to
consider further linear combinations of double parton distributions.
Changing signs of the parameters $a_1 \rightarrow -a_1$, $b_1 \rightarrow
-b_1$, $a_3 \rightarrow -a_3$, $b_3 \rightarrow -b_3$ we get
\begin{align}
  \label{eq:cond2}
  Q_{ab}^- & = c_1 f_{ab} + c_2 \ms h_{a \ms \delta b} 
         + c_3\ms f_{\Delta a \Delta b} - c_4\ms h_{\delta a \ms b} 
         - c_5\ms h_{\delta a \delta b} - c_6\ms h_{\delta a \delta b}^t
  \geq 0 \,.
\end{align}
Adding \eqref{eq:cond1} and \eqref{eq:cond2} gives the simpler inequality
\begin{align}
  \label{eq:cond3}
  B_{ab}^+ = c_1 f_{ab} + c_2\ms h_{a \ms \delta b} 
           + c_3\ms f_{\Delta a \Delta b} \,\geq\, 0 \,,
\end{align}
and interchanging indices ($1 \leftrightarrow 2$ and $3 \leftrightarrow
4$) in the elements of $v$ gives
\begin{align}
  \label{eq:cond4}
  B_{ab}^- = c_1 f_{ab} + c_2\ms h_{a \ms \delta b}
           - c_3\ms f_{\Delta a \Delta b} \,\geq\, 0 \,.
\end{align}
If \eqref{eq:cond1} holds at a given scale for arbitrary values of $a_i$
and $b_i$, then \eqref{eq:cond2} to \eqref{eq:cond4} hold at that scale as
well.

We will see that the evolution equations in the scale $\mu_1$ can be
formulated in terms of $Q_{ab}^+$, $Q_{ab}^-$ and $B_{ab}^-$
alone.\footnote{%
  The combination $B_{ab}^+ = (Q_{ab}^+ + Q_{ab}^-)/2$ is not independent
  and just used as an abbreviation.}
This becomes plausible if we note that these three functions are linear
combinations of $(c_1 f_{ab} + c_2\ms h_{a \ms \delta b})$, $f_{\Delta a
  \Delta b}$ and $(c_4\ms h_{\delta a \ms b} + c_5\ms h_{\delta a \delta
  b} + c_6\ms h_{\delta a \delta b}^t)$ and that for evolution in $\mu_1$
only the polarization of the first parton is relevant but not the
polarization of the second parton.
The linear combinations $Q_{ab}^{\pm}$ may be regarded as generalizations
of the distributions $Q_{\pm} = \half ( q + \bar{q}) \pm \delta q$
introduced in \cite{Bourrely:1997bx}, where it was shown that the Soffer
bound for the quark transversity distribution $\delta q$ is stable under
leading-order evolution to higher scales.

%%%%%%%%%%%%%%

\subsection{Evolution of the linear combinations}
\label{sec:evol-lin-comb}

We now show that the distributions $Q_{ab}^\pm$ and $B_{ab}^\pm$ remain
positive semi-definite under leading-order evolution to higher scales.
This implies the positivity of the spin density matrices and thereby the
validity of the bounds derived in \sect{sec:bounds}.

The evolution equations for the distributions $Q^\pm_{ab}$ are
\begin{align}
\label{eq:QqqEvo}
  \frac{\dd}{\dd\tau_1} 
       \begin{pmatrix} Q_{qb}^+ \\ Q_{qb}^- \end{pmatrix}
  & =  \begin{pmatrix} \delta P_{qq}^+ & \delta P_{qq}^- \\
                       \delta P_{qq}^- & \delta P_{qq}^+ \end{pmatrix} 
       \otimes_1  \begin{pmatrix} Q_{qb}^+ \\ Q_{qb}^- \end{pmatrix} 
               +  \begin{pmatrix} P_{qg}^+ & P_{qg}^- \\
                                  P_{qg}^+ & P_{qg}^- \end{pmatrix} 
       \otimes_1  \begin{pmatrix} B_{gb}^+ \\ B_{gb}^- \end{pmatrix} 
\nonumber\\
  & \quad + \begin{pmatrix} P_{qq}^- & P_{qq}^- \\
                            P_{qq}^- & P_{qq}^- \end{pmatrix}  
       \otimes_1  \begin{pmatrix} B_{qb}^+ \\ B_{qb}^- \end{pmatrix} 
\intertext{for a quark as first parton and}
\label{eq:QgqEvo}
  \frac{\dd}{\dd\tau_1}
       \begin{pmatrix} Q_{gb}^+ \\ Q_{gb}^- \end{pmatrix} 
  & =  \begin{pmatrix} \delta P_{gg}^+ & \delta P_{gg}^- \\
                       \delta P_{gg}^- & \delta P_{gg}^+ \end{pmatrix} 
       \otimes_1  \begin{pmatrix} Q_{gb}^+ \\ Q_{gb}^- \end{pmatrix} 
       + \sum_{a=q, \bar{q}}
         \begin{pmatrix} P_{ga}^+ & P_{ga}^- \\
                         P_{ga}^+ & P_{ga}^- \end{pmatrix} 
       \otimes_1  \begin{pmatrix} B_{ab}^+ \\ B_{ab}^- \end{pmatrix} 
\nonumber\\
  & \quad + \begin{pmatrix} P_{gg}^- & P_{gg}^- \\
                            P_{gg}^- & P_{gg}^- \end{pmatrix}  
       \otimes_1  \begin{pmatrix} B_{gb}^+ \\ B_{gb}^- \end{pmatrix} 
\end{align}
when the first parton is a gluon.  The evolution equations for
$B_{ab}^\pm$ read
\begin{align}
\label{eq:BqqEvo}
  \frac{\dd}{\dd\tau_1} 
       \begin{pmatrix} B_{qb}^+ \\ B_{qb}^- \end{pmatrix} 
  & =  \begin{pmatrix} P_{qq}^+ & P_{qq}^- \\
                       P_{qq}^- & P_{qq}^+ \end{pmatrix} 
       \otimes_1  \begin{pmatrix} B_{qb}^+ \\ B_{qb}^- \end{pmatrix} 
       + \begin{pmatrix} P_{qg}^+ & P_{qg}^- \\
                         P_{qg}^- & P_{qg}^+ \end{pmatrix} 
       \otimes_1  \begin{pmatrix} B_{gb}^+ \\ B_{gb}^- \end{pmatrix} 
\intertext{for a quark and}
\label{eq:BgqEvo}
 \frac{\dd}{\dd\tau_1}
       \begin{pmatrix} B_{gb}^+ \\ B_{gb}^- \end{pmatrix} 
  & =  \begin{pmatrix} P_{gg}^+ & P_{gg}^- \\
                       P_{gg}^- & P_{gg}^+ \end{pmatrix} 
       \otimes_1  \begin{pmatrix} B_{gb}^+ \\ B_{gb}^- \end{pmatrix} 
       + \sum_{a=q, \bar{q}}
         \begin{pmatrix} P_{ga}^+ & P_{ga}^- \\
                         P_{ga}^- & P_{ga}^+ \end{pmatrix} 
       \otimes_1  \begin{pmatrix} B_{ab}^+ \\ B_{ab}^- \end{pmatrix}
\end{align}
for a gluon.  The evolution equations have the same form for antiquarks,
i.e.\ \eqref{eq:QqqEvo} and \eqref{eq:BqqEvo} remain valid if we replace
$q\to \bar{q}$ everywhere (except in the label $b$ for the second parton,
which always remains fixed when we consider evolution in $\mu_1$).

The splitting functions appearing in the above equations are defined as
\begin{align}
  P_{ab}^\pm & = \frac{1}{2}\ms
    \bigl( P_{ab} \pm P_{\Delta a \Delta b} \bigr) \,,
&
  \delta P_{ab}^\pm & = \frac{1}{2}\ms
    \bigl( P_{\Delta a \Delta b} \pm P_{\delta a \delta b} \bigr)
\end{align}
for all parton indices $a$ and $b$.  We remark that the kernels $P_{ab}^+$
($P_{ab}^-$) correspond to the case where the parton helicity is conserved
(flipped).
The only splitting functions that receive contributions from virtual
graphs and hence contain a plus-prescription or an explicit $\delta$
function are
\begin{align}
  P_{qq}^+ & = \frac{C_F}{2}\, \biggl[\, \frac{2\ms (1+z^2)}{(1-z)_+} 
                      + 3\ms \delta(1-z) \biggr] \,,
\nonumber \\
  \delta P_{qq}^+ & = \frac{C_F}{2}\, \biggl[\, \frac{(1+z)^2}{(1-z)_+} 
                      + 3\ms \delta(1-z) \biggr] \,,
\nonumber\\
  P_{gg}^+ & =  2 N_c\, \biggl[\, \frac{z}{(1-z)_+}
                        + \frac{(1-z) (1+z)^2}{2 z} \,\biggr]
                        + \frac{\beta_0}{2}\, \delta(1-z) \,,
\nonumber\\
  \delta P_{gg}^+ & = 2 N_c\, \biggl[\, \frac{z}{(1-z)_+}
                        + 1-z \,\biggr]
                        + \frac{\beta_0}{2}\, \delta(1-z)
\end{align}
with $N_c = 3$, $C_F = 4/3$ and
\begin{align}
\beta_0 = \frac{11}{3}\ms N_c - \frac{2}{3}\ms n_f ,
\end{align}
where $n_f$ is the number of active quark flavors.  
They are all positive for $0<z<1$ but have negative contributions at $z=1$
that arise from the plus-prescription, whose form is recalled in
\eqref{eq:def-plus-presc}.  In \app{ap:pos} we show explicitly that the
virtual contribution to evolution cannot change the sign of the
distributions, which has previously been argued to be the case based on
the probabilistic interpretation of leading-order evolution and its
relation to the Boltzmann equation
\cite{Durand:1986te,Collins:1988wj,Bourrely:1997bx}.  The reason for this
property is that the virtual contribution to the evolution of a function
is proportional to the function itself.  We can then conclude that the
diagonal terms in the evolution equations \eqref{eq:QqqEvo} to
\eqref{eq:BgqEvo} preserve positivity.  The off-diagonal kernels
\begin{align}
  P_{qq}^- & = 0 \,,
&
  P_{gg}^- & =  N_c\ms (1-z)^3  \big/ z \,,
\nonumber\\
  \delta P_{qq}^- & = C_F\ms (1-z) \big/ 2 \,,
&
  \delta P_{gg}^- & = 2 N_c\ms (1-z)
\intertext{and}
  P_{qg}^+ & =  z^2 \big/ 2 \,,
&
  P_{gq}^+ & =  C_F \big/ z \,,
\nonumber\\
  P_{qg}^- & =  (1-z)^2 \big/ 2 \,,
&
  P_{gq}^- & =  C_F\ms (1-z)^2 \big/ z \,.
\end{align}
are all positive or zero for $0<z<1$ and regular at $z=1$.  Therefore
they only reinforce positivity.  In summary, if we have positive
semi-definite initial conditions for all functions $Q_{ab}^\pm$ and
$B_{ab}^\pm$ at some scale, then evolution to higher scales preserves this
property.  A more explicit derivation is given in \app{ap:pos}.

%%%%%%%%%%%%%%%%%%%%%%%%%%%%%%%%%%%%%%%%%%%%%%%%%%%%%%%%%%%%%%

\section{Conclusions}
\label{sec:sum}

We have derived spin density matrices for double parton distributions of
quarks, anti-quarks and gluons.  These matrices reveal the full
polarization structure of two partons in an unpolarized proton and show
the correspondence between the different polarized double parton
distributions and parton helicities.  The probabilistic interpretation of
the double parton distribution for an arbitrary polarization state of the
two partons gives upper limits on the size of spin correlations.  These
positivity bounds can be useful for modeling the otherwise poorly
constrained double parton distributions and for deriving upper limits on
spin effects in double hard scattering processes.  We have shown that the
homogeneous leading-order evolution equations preserve the bounds when
going from lower to higher scales.

%%%%%%%%%%%%%%%%%%%%%%%%%%%%%%%%%%%%%%%%%%%%%%%%%%%%%%%%%%%%%%

\appendix

%%%%%%%%%%%%%%%%%%%%%%%%%%%%%%%%%%%%%%%%%%%%%%%%%%%%%%%%%%%%%%

\section{Evolution equations and splitting functions}
\label{ap:evo}

For completeness we give here the leading-order evolution equations for
the first parton in the double parton distributions.  When the first
parton is a quark, we have
\begin{align}
  \frac{\dd f_{qb}}{\dd\tau_1}
    & = P_{qq} \otimes_1 f_{qb} + P_{qg} \otimes_1 f_{gb} \,,
\nonumber \\
  \frac{\dd f_{q \ms \delta b}}{\dd\tau_1}
    & =  P_{qq} \otimes_1 f_{q \ms \delta b}
       + P_{qg} \otimes_1 f_{g \ms \delta b} \,,
\nonumber \\
  \frac{\dd f_{\Delta q \Delta b}}{\dd\tau_1}
    & = P_{\Delta q \Delta q} \otimes_1 f_{\Delta q \Delta b}
      + P_{\Delta q \Delta g} \otimes_1 f_{\Delta g \Delta b} \,,
\nonumber \\
  \frac{\dd f_{\delta q \ms b}}{\dd\tau_1}
    & = P_{\delta q \delta q} \otimes_1 f_{\delta q \ms b} \,,
\quad\;
  \frac{\dd f_{\delta q \delta b}}{\dd\tau_1}
      = P_{\delta q \delta q} \otimes_1 f_{\delta q \delta b} \,,
\quad\;
  \frac{\dd f_{\delta q \delta b}^t}{\dd\tau_1}
      = P_{\delta q \delta q} \otimes_1 f_{\delta q \delta b}^t
\end{align}
for $b = q,\bar{q},g$.  The arguments of the distributions are as in
\eqref{eq:evol-parton-1} and \eqref{eq:otim}.  Analogous equations hold if
the first parton is an antiquark.  For gluons we have
\begin{align}
  \frac{\dd f_{gb}}{\dd\tau_1}
    & = P_{gg} \otimes_1 f_{gb}
      + \sum_{a=q, \bar{q}} P_{ga} \otimes_1 f_{ab} \,,
\nonumber \\
  \frac{\dd f_{g \ms \delta b}}{\dd\tau_1}
    & =  P_{gg} \otimes_1 f_{g \ms \delta b}
      + \sum_{a=q, \bar{q}} P_{ga} \otimes_1 f_{a \ms \delta b} \,,
\nonumber \\
  \frac{\dd f_{\Delta g \Delta b}}{\dd\tau_1}
    & = P_{\Delta g \Delta g} \otimes_1 f_{\Delta g \Delta b} 
      + \sum_{a=q, \bar{q}} 
        P_{\Delta g \Delta a} \otimes_1 f_{\Delta a \Delta b} \,,
\nonumber \\
  \frac{\dd f_{\delta g \ms b}}{\dd\tau_1}
    & = P_{\delta g \delta g} \otimes_1 f_{\delta g \ms b} \,,
\quad\;
  \frac{\dd f_{\delta g \delta b}}{\dd\tau_1}
      = P_{\delta g \delta g} \otimes_1 f_{\delta g \delta b} \,,
\quad\;
  \frac{\dd f_{\delta g \delta b}^t}{\dd\tau_1}
      = P_{\delta g \delta g} \otimes_1 f_{\delta g \delta b}^t \,.
\end{align}
The leading-order splitting functions have been derived in
\cite{Altarelli:1977zs,Artru:1989zv}.   They
are given by
\begin{align}
  \label{eq:quark-kernels}
  P_{qq}(z) & =  C_F\, \biggl[ \frac{1+z^2}{(1-z)_+}
                 + \frac{3}{2}\, \delta(1-z) \biggr] \,,
\nonumber \\
  P_{\Delta q \Delta q}(z) & = P_{qq}(z) \,,
\nonumber \\
  P_{\delta q \delta q}(z) & = P_{qq}(z) -  C_F (1-z)
  \phantom{\frac{1}{1}}
\end{align}
for quark-quark transitions and by
\begin{align}
  \label{eq:gluon-kernels}
  P_{gg}(z) & =  2 N_c\, \biggl[\, \frac{z}{(1-z)_+}
                 + \frac{(1-z)(1+z^2)}{z} \,\biggr]
                 + \frac{\beta_0}{2}\, \delta(1-z) \,,
\nonumber\\
  P_{\Delta g \Delta g}(z) & = P_{gg}(z) - 2 N_c\, \frac{(1-z)^3}{z} \,,
\nonumber\\
  P_{\delta g \delta g}(z) & = P_{gg}(z)
                               - 2 N_c\, \frac{(1-z) (1+z^2)}{z}
\end{align}
for gluons.  The splitting functions that mix quarks and gluons read
\begin{align}
  \label{eq:mixing-kernels}
  P_{qg} & =  \frac{z^2 + (1-z)^2}{2} \,,
&
  P_{gq} & =  C_F\, \frac{1+ (1-z)^2}{z} \,,
\nonumber\\
  P_{\Delta q \Delta g} & =  \frac{z^2 - (1-z)^2}{2} \,,
&
  P_{\Delta g \Delta q} & =  C_F\, \frac{1-(1-z)^2}{z} \,.
\end{align}
As already mentioned below \eqref{eq:otim}, the splitting functions are
identical for quarks and antiquarks, i.e.\ \eqref{eq:quark-kernels} and
\eqref{eq:mixing-kernels} remain valid if we replace $q\to \bar{q}$.  At
leading order in $\alpha_s$ there are no direct transitions between quarks
and antiquarks.

%%%%%%%%%%%%%%%%%%%%%%%%%%%%%%%%%%%%%%%%%%%%%%%%%%%%%%%%%%%%%%

\section{Elements of a stability proof}
\label{ap:pos}

In this appendix we show in more detail that the evolution equations in
\sect{sec:evol-lin-comb} preserve positivity, taking particular care of
the negative terms in the splitting functions that arise from virtual
graphs and are implicit in the plus-prescription.  We first consider the
evolution of a single distribution and then extend the argument to the
full coupled system of evolution equations.

We examine a function evolving as
\begin{align}
  \label{eq:original-evolution}
  \frac{\dd}{\dd\tau}\, f(x,\tau) & = \int_{x}^{v} \frac{d u}{u}\,
     P\left( \frac{x}{u} \right) f(u,\tau)
\end{align}
with $0< x < v \leq 1$ and separate the splitting function as
\begin{align}
  P(z) = \frac{P_s(z)}{(1-z)_+} + P_r(z) + P_\delta\, \delta(1-z) \,,
\end{align}
where $P_s(z)$ and $P_r(z)$ are positive semi-definite for $0<z<1$ and
regular at $z=1$.  The constant $P_\delta$ may be positive, negative or
zero.  The plus-prescription is defined as usual by
\begin{align}
  \label{eq:def-plus-presc}
  \left[ s(z) \right]_+ = s(z) - \delta(1-z)\int_0^1 dz'\, s(z') \,,
\end{align}
where it is understood that the non-integrable singularity in the last
term cancels when \eqref{eq:def-plus-presc} is integrated over with a
smooth test function.
The plus-prescription part of the convolution in
\eqref{eq:original-evolution} can be written as
\begin{align}
& \int_{x}^{v} \frac{d u}{u} \frac{P_s(x/u)}{(1 - x/u)_+}\, f(u,\tau)
\nonumber \\[0.2em]
 & \qquad
 = \int_{x+\epsilon}^{v} \!d u\; \frac{P_s(x/u)}{u - x}\, f(u,\tau)
   + \int_0^{x-\epsilon} \!d u\; \frac{P_s(1)}{u - x}\, f(x,\tau)
   + \mathcal{O}(\epsilon) \,,
\end{align}
where for the error estimate we have assumed that $f(u,\tau)$ is
differentiable at $u=x$.  Defining
\begin{align}
  \label{eq:gh-def}
  g_{\epsilon}(x,\tau; f) &= \int_{x+\epsilon}^{v} \!d u\;
      \biggl[\ms \frac{P_s(x/u)}{u - x} +
                 \frac{P_r(x/u)}{u} \ms\biggr]\, f(u,\tau) \,,
\nonumber \\[0.2em]
  h_{\epsilon}(x) &= {}- P_\delta + P_s(1)
      \int_0^{x-\epsilon} \! \frac{d u}{x - u}
\end{align}
we can approximate the evolution of $f$ by
\begin{align}
  \label{eq:epsilon-evolution}
  \frac{\dd}{\dd\tau}\, f(x,\tau) =
     g_{\epsilon}(x,\tau; f)  - h_{\epsilon}(x)\, f(x,\tau)
\end{align}
with an error that becomes arbitrarily small for $\epsilon \to 0$.  In a
more formal proof, one would replace $f$ with $f_{\epsilon}$ in
\eqref{eq:epsilon-evolution} and show that $\lim\limits_{\epsilon\to 0}
f_{\epsilon}$ is a solution of \eqref{eq:original-evolution} .  We now
rewrite \eqref{eq:epsilon-evolution} as
\begin{align}
\label{eq:intf}
\frac{\dd}{\dd\tau} \left[ e^{\tau h_{\epsilon}(x)} f(x,\tau) \right]
  = e^{\tau h_{\epsilon}(x)}\, g_{\epsilon}(x,\tau; f) \,.
\end{align}
Since $g_{\epsilon}$ is the convolution of $f(x,\tau)$ with a positive
semi-definite integral kernel, the r.h.s.\ of this equation is positive
semi-definite as long as $f(x,\tau)$ is.  With initial conditions
$f(x,\tau_0) \ge 0$ for all $x$ at a starting scale $\tau_0$, the function
$e^{\tau h_{\epsilon}(x)} f(x,\tau)$ can therefore not decrease as $\tau$
increases, so that $f(x,\tau)$ stays positive semi-definite for all $\tau
> \tau_0$.  We note that the sign of $h_{\epsilon}(x)$ and thus of the
constant $P_\delta$ is irrelevant for this argument.

We now consider the coupled system of evolution equations given by
\eqref{eq:QqqEvo} to \eqref{eq:BgqEvo}.  Using a vector notation
$f^i(x,\tau)$ for the $8 n_f + 4$ functions $Q^+_{ab}, Q^-_{ab}, B^+_{ab},
B^-_{ab}$ with $a=q, \bar{q}, g$ (and $b$ fixed), we can cast their
evolution into the form
\begin{align}
  \label{eq:coupled-evolution}
\frac{\dd}{\dd\tau}\, f^i(x,\tau) &=
  g^i_{\epsilon}(x,\tau; f^i) - h^i_{\epsilon}(x)\, f^i(x,\tau) 
  + \sum_{i\neq j} \int_x^v \frac{d u}{u}\, 
        P^{ij}\left( \frac{x}{u} \right) f^j(u,\tau)
\end{align}
with $i=1, \ldots, 8 n_f + 4$.  Here $g^i_{\epsilon}$ and $h^i_{\epsilon}$
are defined as in \eqref{eq:gh-def} with regular and positive
semi-definite functions $P^i_s(z)$ and $P^i_r(z)$.  The mixing kernels
$P^{ij}(z)$ in \eqref{eq:coupled-evolution} are regular and positive
semi-definite as well.  Rewriting the evolution as
\begin{align}
\frac{\dd}{\dd\tau}
  \left[ e^{\tau h_{\epsilon}(x)} f^i(x,\tau) \right]
  & = e^{\tau h_{\epsilon}(x)}
    \Biggl[ \ms g^i_{\epsilon}(x,\tau; f^i)
            + \sum_{i\neq j} \int_x^v \frac{d u}{u}\, 
              P^{ij}\left( \frac{x}{u} \right) f^j(u,\tau) \Biggr]
\end{align}
we see that if one has initial conditions $f^j(x,\tau_0) \ge 0$ for all
$j$ then all functions $f^j(x,\tau)$ remain positive semi-definite for
$\tau > \tau_0$.

%%%%%%%%%%%%%%%%%%%%%%%%%%%%%%%%%%%%%%%%%%%%%%%%%%%%%%%%%%%%%%

% the following lines create an entry in the table of contents
\phantomsection

\end{document}